\documentclass[]{aastex63}

\usepackage{amsmath}
\usepackage{graphicx}
\usepackage{float}
\usepackage{tabularx}
\usepackage{url}
\usepackage{subfigure}

\usepackage{chemformula}
\newcommand{\ce}[1]{\ch{#1}}
\shorttitle{Outer Edge of the Venus Zone}
\shortauthors{Vidaurri et al.}

\begin{document}
\title{The Outer Edge of the Venus Zone Around Main-Sequence Stars}

\author[0000-0002-5042-785X]{Monica R. Vidaurri}
\affiliation{Department of Physics and Astronomy, Howard University, Washington, DC}
\affiliation{NASA Goddard Space Flight Center, Greenbelt, MD}
\affiliation{Center for Research and Exploration in Space Science and Technology, NASA/GSFC, Greenbelt, MD}
\affiliation{Department of Geological Sciences, Stanford University, Stanford, CA}

\author[0000-0003-2052-3442]{Sandra T. Bastelberger}
\affiliation{University of Maryland, College Park, MD}
\affiliation{NASA Goddard Space Flight Center, Greenbelt, MD}
\affiliation{Center for Research and Exploration in Space Science and Technology, NASA/GSFC, Greenbelt, MD}
\affiliation{NASA GSFC Sellers Exoplanet Environments Collaboration, Greenbelt, MD}

\author[0000-0002-7188-1648]{Eric T. Wolf}
\affiliation{Laboratory for Atmospheric and Space Physics, Department of Atmospheric and Oceanic Sciences, University of Colorado, Boulder, Boulder, CO} 
\affiliation{NASA GSFC Sellers Exoplanet Environments Collaboration, Greenbelt, MD}
\affiliation{NExSS Virtual Planetary Laboratory, Seattle, WA}f

\author[0000-0003-0354-9325]{Shawn Domagal-Goldman}
\affiliation{NASA Goddard Space Flight Center, Greenbelt, MD}
\affiliation{NExSS Virtual Planetary Laboratory, Seattle, WA}
\affiliation{NASA GSFC Sellers Exoplanet Environments Collaboration, Greenbelt, MD}

\author[0000-0002-5893-2471]{Ravi Kumar Kopparapu}
\affiliation{NASA Goddard Space Flight Center, Greenbelt, MD}
\affiliation{NASA GSFC Sellers Exoplanet Environments Collaboration, Greenbelt, MD}

\begin{abstract}

A key item of interest for planetary scientists and astronomers is the habitable zone: the distance from a host star where a terrestrial planet can maintain necessary temperatures in order to retain liquid water on its surface. However, when observing a system's habitable zone, it is possible that one may instead observe a Venus-like planet. We define "Venus-like" as greenhouse-gas-dominated atmosphere occurring when incoming solar radiation exceeds infrared radiation emitted from the planet at the top of the atmosphere, resulting in a runaway greenhouse. Our definition of Venus-like includes both incipient and post-runaway greenhouse states. Both the possibility of observing a Venus-like world and the possibility that Venus could represent an end-state of evolution for habitable worlds, requires an improved understanding of the Venus-like planet; specifically, the distances where these planets can exist. Understanding this helps us define a "Venus zone" - the region in which Venus-like planets could exist - and assess the overlap with the aforementioned ``Habitable Zone''. In this study, we use a 1D radiative-convective climate model to determine the outer edge of the Venus zone for F0V, G2V, K5V, and M3V and M5V stellar spectral types. Our results show that the outer edge of the Venus zone resides at 3.01, 1.36, 0.68, 0.23, and 0.1 AU, respectively. These correspond to incident stellar fluxes of 0.8, 0.55, 0.38, 0.32, and 0.3 S$_\odot$, respectively, where stellar flux is relative to Earth (1.0). These results indicate that there may be considerable overlap between the habitable zone and the Venus zone.

\end{abstract}

\section{Introduction} \label{sec:intro}
\par
Exoplanet observation and detection have drastically improved since the discovery of the first exoplanet with a pulsar host star three decades ago \citep{wolszczan_planetary_1992}, and the first exoplanet discovery around a main-sequence star \citep{mayor_jupiter-mass_1995}. Titans of exoplanet observation technology such as the Kepler and K2, Spitzer, Hubble, and TESS missions – discoveries dominated by the transit method of detection – have drastically improved technological capabilities of both finding and characterizing these exoplanets in various system environments \citep{charbonneau_when_2007, Borucki977, batalha_exploring_2014, howell_k2_2014, barclay_revised_2018, deming_highlights_2020, kane_science_2020}. In addition to greater precision of detection and classification which resulted in radical increases to the number of exoplanets observed, improvements to exoplanet observation include more precise transiting follow-up capabilities for planets discovered via radial velocity technique  \citep{dalba_predicted_2019}, improved occurrence rates for planet types of interest such as potentially habitable planets \citep{dressing_occurrence_2015}, refined study of atmospheric composition via transmission spectroscopy \citep{Kempton_2018}, more advanced detection of carbon-based molecules in planetary atmospheres \citep{deming_highlights_2020}, and many others. And we are awaiting more advanced discoveries from future space-based missions such as JWST, Large UV/Optical/IR Surveyor (LUVOIR),  Habitable Exoplanet Observatory (HabEx), Origins, and Large InterferometeR For Exoplanets (LIFE) mission concepts, and ground-based observatories such as the Extremely Large Telescope (ELT), Very Large Telecsope (VLT), and Giant Magellan Telescope (GMT). These modern and next generation observatories will permit the efficient and effective study of the topics mentioned above, singular planets of interest, and will allow us to further define areas of interest within a given system, such as the habitable zone \citep{gardner_james_2006,rodler_feasibility_2014, bolcar_large_2017,cooray_origins_2018,serindag_testing_2019, gaudi_habitable_2020, angerhausen_large_2021}. Thus, the observation of terrestrial planets ($\lessapprox$ 2R$_\oplus$) \citep{lammer_origin_2014,lopez_understanding_2014} residing within the habitable zone are of immense interest. However, evidence has shown that terrestrial planets detected by transiting observations could possibly be Venus-like planets \citep{kane_frequency_2014}. Here, we define "Venus-like" as a state which includes both incipient and post-runaway greenhouse states. In past model experiments, the inner edge of the habitable zone was defined as the region where incipient runaway conditions will occur, given modern Earth's atmospheric composition \citep{Kopparapu2013, Kopparapu2014}. This happens when incoming solar radiation is predicted to exceed infrared radiation emitted from the planet at the top of the atmosphere, resulting in a runaway greenhouse state. This definition of an incipient greenhouse assumes an evolution into a post-runaway greenhouse state, when water has been lost to space. We consider planets in this case to be "Venus-like." A planet would avoid the fate of being Venus-like (either incipient or post-runaway greenhouse) when an excess of the greenhouse gas dominating the atmosphere begins condensing, signaling maximum greenhouse gas warming has been reached and cooling and atmospheric collapse are commencing. It is worth noting that a detection of Venus-like planets will likely be close their host star and therefore relatively easier to detect, due to a preference for shorter orbital distances by transiting observations \citep{kane_constraining_2008}.\par
In addition to the likelihood of observing a Venus-like planet within a given habitable zone, Venus-like planets, and subsequently the Venus zone, might be crucial to understanding the evolution of Earth analogs themselves. Venus shares many similarities with Earth in terms of mass, radius, overall bulk composition \citep{goettel_density_1982,kaula_tectonics_1994,svedhem_venus_2007}, and is believed to have had an Earth-like environment in the past \citep{way_was_2016, khawja_tesserae_2020,way_venusian_2020}, although it is also hypothesized that Venus could have been like it is today since its beginnings \citep{hamano_emergence_2013, turbet_daynight_2021}. Venus' evolution into a post-runaway greenhouse state may suggest that the Venus-like planets that we observe today may have been Earth-like, or akin to modern Earth conditions, in the past, or that a runaway greenhouse may one day be the fate of our Earth \citep{ingersoll_runaway_1969,lapotre_probing_2020}. Kopprapu et al. \citeyear{Kopparapu2013} and \citeyear{Kopparapu2014} have previously defined incident stellar fluxes that allow for the runaway greenhouse to occur, giving us nominal estimates of a Venus zone: the boundaries of incident stellar flux received from a host star which allow a terrestrial atmospheric environment to enter a runaway greenhouse state, and where observers can expect to find Venus-like planets. These prior studies focused the distances at which runaway greenhouses become likely. However, these studies did not investigate how greenhouse gas abundance (in this case, \ce{CO2}) could impact that range. In other words, prior work did not set an ``outer limit" to the region around a star for which incipient runaway conditions - and therefore Venus-like worlds - could occur. For this work, we examine the distance in which a planet can enter runaway greenhouse conditions. We provide a theoretical outer edge of the Venus zone for F, G, K, and M stars, including the warming effects from \ce{CO2}-rich atmospheres. We also visualize the distance into a system's habitable zone where it is entirely possible to observe either incipient or post-runaway greenhouse atmospheres; Venus-like planets.\par

\section{Model and Parameter Description} \label{sec:model}
 \par We use the climate portion of the 1D photochemical-climate model \texttt{Atmos} for this project. The climate portion of this model was initially developed by \cite{kasting_climatic_1986}, and was recently updated in \cite{Kopparapu2014, Ben2020}. Calculations completed in this model use six solar zenith angles centered around 60$^\circ$, Earth radius, and Earth gravity. The model surface bond albedo is set to 0.24, which reproduces modern Earth's global average surface temperature of $\sim$ 288 K. This value is higher than Earth's surface albedo of $\sim$0.125 \citep{Trenberth2009} but lower than current estimates of the planetary bond albedo (including cloud cover) of $\sim$0.3 \citep{stephens_albedo_2015}, to compensate for the lack of clouds in the model. In addition, the model uses updated \ce{H2O} and \ce{CO2} $k$-coefficients, which are central to climate modeling. These coefficients are used in the radiative transfer calculation to describe the amount of energy that is absorbed by a given species in each layer of the planetary atmosphere. A detailed justification of the choice of the surface bond albedo value, as well as a description of these updated $k$-coefficients, are provided in Section \ref{sec:alb} and Section \ref{sec:rt}. \par
 
% In the climate model's input of partial pressures for each constituent present in the atmosphere, we do not include the partial pressure of water ($p_{\textnormal{H\textsubscript{2}O}}$) in the total atmospheric pressure $p_{tot}$; each species is "dry". 
 \par
 For clarification purposes of the future use of ``convergence" of this model, the model is considered converged when the divergence of the flux at the top of the atmosphere (TOA) is minimal ($\sim$ 10$^{-3}$).\par

\subsection{Updated k-Coefficients}\label{sec:rt}

New $k$-coefficients for \ce{H2O} and \ce{CO2} were calculated using HELIOS-K\footnote{ \url{https://github.com/exoclime/HELIOS-K}}, an ultrafast GPU-driven correlated-$k$ sorting program \citep{Grimm_2015}. For \ce{H2O} we use the HITRAN2016 line-list \citep{GORDON20173}, assuming 25 cm$^{-1}$ line cut-offs using Lorentz profiles and with the plinth\footnote{Plinth refers to the subtracted line-center value from the overlayed continuum, such that the line center is not counted twice when both the HITRAN and continuum lines are re-added} removed. For \ce{CO2} we also use the HITRAN2016 database, but we assume 500 cm$^{-1}$ line cut-offs using the Perrin and Hartman sub-Lorentzian line profiles \citep{Perrin_1989}. Overlapping absorption from multiple gas species is treated assuming that the gases are uncorrelated \citep{shi:2009}.  Continuum absorption for \ce{H2O} and \ce{CO2} are included separately from the correlated $k$-coefficients as follows. For \ce{H2O} we include the foreign and self-broadened continuum coefficients using the BPS formalism \citep{Paynter2011}.  For \ce{CO2} we include \ce{CO2}-\ce{CO2} collision induced absorption following \citet{wordsworth:2010}.  These conventions represent the current standard practices for the treatment of \ce{H2O} and \ce{CO2} lines within coarse spectral resolution climate model radiation schemes for planetary atmospheres.  Here we have chosen to use the BPS continuum, because it is derived from laboratory measurements taken at higher temperatures compared to the commonly used MT-CKD continuum.

\subsection{Surface Bond Albedo}\label{sec:alb}
Updating $k$-coefficients in our model also requires that the surface albedo be re-tuned. As noted in Section \ref{sec:model}, this tuning results in an albedo of 0.24, higher than current surface albedo estimates of $\sim$0.125 and lower than current estimates of average albedo accounting for cloud coverage of $\sim$0.3.\par

The version of \texttt{Atmos} we used does not include cloud prescription (see \cite{Thomas2018} for a version of the model that includes clouds). Instead, it relies on the input of surface albedo in order to simulate the effects of clouds. However, clouds have complex altitude-dependent effects on radiative transfer. For example, clouds residing lower in the atmosphere reflect sunlight back into space, in turn raising the albedo of the planet, resulting in a cooling effect. Clouds residing in the upper atmosphere (cirrus clouds, for example) warm the surface by retaining heat absorbed from solar radiation, albeit with a much smaller effect on albedo. Our model has one albedo value - applied at the surface - that needs to account for these competing effects. Thus, a qualitatively consistent tuning would be represented by an albedo in between the Earth's surface albedo and it's upper atmosphere albedo. While a previous surface albedo value of 0.32 was used in conjunction with an earlier set of $k$-coefficients \cite{Kopparapu2013}, we contend that this albedo would disproportionately represent the cooling effect of lower-atmosphere clouds and deny the warming effects of upper-atmosphere clouds.

\subsection{The \ce{CO2}, \ce{N2}, and \ce{H2O} Atmosphere}\label{sec:atmo}
The atmospheric constituents included in the model are gaseous Ar, CH$_4$, C$_2$H$_6$, CO$_2$, N$_2$, O$_2$, H$_2$, and NO$_2$. The model accounts for \ce{Ar} and \ce{C2H6} mainly in terms of the scale height calculations, which maps pressure to altitude, and does not assume these to be spectrally active in radiative transfer.  The code can also model the radiative effect of hydrocarbon aerosols, but no such aerosols were included in our simulations. We simulated atmospheres consisting of spectrally active \ce{CO2}, \ce{N2}, and \ce{H2O}; representing the major constituents of a terrestrial \ce{CO2}-dominated incipient greenhouse atmosphere. Although we directly adjust partial pressures for \ce{CO2} and \ce{N2} in the model parameters, we do not set a partial pressure for water. Instead, it is computed by the model and added to the non-water components to calculate the total pressure. Our model uses the Manabe and Whetherald parameterization \citep{ThermalEquilibriumoftheAtmospherewithaGivenDistributionofRelativeHumidity}, which determines the partial pressure of \ce{H2O}, $p_{\textnormal{H\textsubscript{2}O}}$ by assuming an Earth-like relative humidity profile in the troposphere. \par

A \ce{CO2}-\ce{N2}-\ce{H2O} atmosphere may exist in the habitable zones of stars for various reasons. For M-type stars, the high luminosity pre-main sequence phase can remove water from a terrestrial atmosphere \citep{luger_extreme_2015}. High radiation drives photolysis of \ce{H2O} as well as loss of \ce{H} into space. This process could eventually result in a \ce{CO2}-rich atmosphere via outgassing and an exo-Venus residing within the habitable zone. For K, G, and F stars, carbon-silicate cycling effectively acts as a ``thermostat'' to counteract excessive warming facilitated by \ce{CO2} \citep{walker_negative_1981, menou_climate_2015, haqq-misra_limit_2016} and may occur more frequently than believed depending on the planetary interior \citep{kadoya_conditions_2014}. These stabilizing feedbacks would also expand the outer edge of any stable climate regime \citep{kasting_evolution_2003}. For these reasons, \ce{CO2}-\ce{N2}-\ce{H2O} atmospheres were amongst the first to be studied for defining the limits of the habitable zone, and in particular for defining its outer edge \citep{Kopparapu2013,Kopparapu2014}; here we are conducting a parallel exploration of the outer limits of the Venus zone, assuming a planet with a source of volatiles and carbonate-silicate feedbacks. However, it is unknown how common this cycling is on temperate terrestrial planets. In other words, there is no universal mechanism preventing the buildup of \ce{CO2} in terrestrial atmospheres around these stars. Even if the carbon-silicate cycle is common in terrestrial planets, it could potentially vary widely depending on the mass and interior structure of the planet itself. For example, it could be that a planet's mass could change the rate of mantle convection, and therefore impact plate tectonics \citep{oneill_geological_2007, valencia_inevitability_2007}. The cooling rate of a planet, and therefore its crustal and tectonic properties, can also be linked to a planet's size \citep{seales_note_2021}. When we consider the absence of efficient carbon-silicate cycling, \ce{CO2} may accumulate in an atmosphere \citep{haqq-misra_limit_2016}. \par
Finally, we rely on CO$_2$ dominated atmospheres, whether caused by post-runaway greenhouse conditions or the general failure of carbon-silicate cycling, to show the maximum possible distance from a host star where a Venus-like planet may occur. We further discuss this approach and the ways that we can more accurately define the Venus zone in Section \ref{sec:results}. \par 

\section{Methodology}
\label{sec:fwd}
First, we use the model to find our ``Venus analog" by starting with present Earth values for CO$_2$ and N$_2$ mixing ratios, incident stellar flux, surface pressure, and temperature. We then steadily increase CO$_2$ partial pressure until the model experiences a runaway greenhouse, ultimately crashing the model. Note that a runaway greenhouse is achieved when excessive atmospheric water vapor effectively closes the radiative window to space, hopelessly trapping thermal radiation and allowing incoming solar radiation to exceed infrared radiation emitted from the planet at the top of the atmosphere even as the surface temperature continues to climb \citep{goldblatt_runaway_2012}. The \texttt{Atmos} model cannot properly handle these conditions, so we use the parameters of the last stable solution the model provides just prior to the crash. These parameters become our "Venus analog", representing a transition into an incipient greenhouse state, and assuming the analog will evolve into a post-runaway greenhouse. Next, our Venus analog parameters (as described at the end of Section \ref{sec:fwd}) are kept constant, while we steadily decrease stellar flux until CO$_2$ condenses in the atmosphere. In our definition of a Venus-like planet (stated in Section \ref{sec:intro}), a planet relinquishes its Venus-like conditions when an excess of the greenhouse gas dominating the atmosphere (\ce{CO2} for this work) begins condensing, meaning that maximum greenhouse gas warming has been achieved. By holding our analog parameters constant while decreasing the stellar flux, we are effectively probing the distance from a given star in which these Venus-like conditions can no longer be maintained and a dense \ce{CO2} may begin condensing and ultimately collapsing.

We utilize this model for an F0V star, Sun-like G2V star, K5V star, and two different M star templates – M3V and M5V. We chose these spectral classes as representatives for the F, G, K, and M spectral types, as we wanted to pursue this model using these types due to their feasibility for observational follow-up. We include two different M star temperatures due to the feasibility of study by ground- and space-based observation presented by M star systems. Not only are these planets easier to detect due to shorter orbital periods, but present higher signal-to-noise ratios due to increased number of transits compared to larger, warmer stars \citep{kopparapu_habitable_2017}, resulting in piqued interest in M star systems. Our G star template represents our sun, with a temperature of 5780 K. The F star template is 7200 K, our K star template is 4400 K, we use a 3000 K and 3400 K M star template. Each of these templates assumes solar metallicity, where the relative abundance of iron to hydrogen is sun-like, i.e. [Fe/H] is zero \citep{Kopparapu2013}. 

We use the ``BT\_Settl" grid of models to simulate the spectral types used in our calculations \citep{allard_model_2003,allard_kh2_2007}. These models cover a range of stellar temperatures from 2600 to 7200 K, as well as metallicities [Fe/H] from -4.0 to +0.5. Section 4.1 in \cite{Kopparapu2013} compares the BT\_Settl models to IRTF and CRIRES data, and shows that the BT\_Settl models are sufficient in reproducing spectral features of each star type.

We begin with present Earth partial pressures of \ce{CO2}  and \ce{N2}, corresponding to mixing ratios of $f_{\textnormal{CO}_2}$ = 3.3$\times 10^{-04}$, and $f_{\textnormal{N}_2}$ = 0.78. From here, we continually increase the partial pressure of CO$_2$ ($p_{\textnormal{CO}_2}$) while keeping $p_{\textnormal{N}_2}$ constant . Each time $p_{\textnormal{CO}_2}$ is increased, we adjust total atmospheric pressure ($p_{tot}$), which is the sum of pressures exerted on the planet by each individual gas accounted for in the model. Note that while \ce{CO2} and \ce{N2} are directly adjusted, \ce{H2O} is also spectrally active in our model and calculated by the model itself, so $p_{tot}$ consists of the sum of $p_{\textnormal{CO}_2}$, $p_{\textnormal{N}_2}$, and $p_{\textnormal{H}_2O}$.

From a present Earth analog, we begin our first iteration by doubling CO$_2$ from $p_{CO2}$ = 3.3$\times 10^{-04}$ bar to $p_{CO2}$ = 6.6$\times 10^{-04}$ bar, adding this new partial pressure to our Earth-like N$_2$ partial pressure, and using these two constituents to calculate $p_{totnew}$. After finding $p_{totnew}$, the mixing ratios (interchangeable in the model with partial pressures) of the rest of the species present can be found; for our case, the new mixing ratio of N$_2$ is calculated. For each subsequent iteration, a new CO$_2$ is established, and the values of each species set by the previous iteration are used, until we achieve our last stable run that produces a converged solution. \par
As a result, our Venus analog with a Sun-like star (5800 K) is an 8.1 bar atmosphere represented by 90\% CO$_2$ and 10\% N$_2$, with a surface temperature of 376 K. For a 7200 K F-type star, this analog has an 11 bar surface pressure with 93\% CO$_2$ and 7\% N$_2$, and a surface temperature of 373 K. A Venus analog with a 4400 K K-type star is characterized with a 4.9 bar surface pressure, made up of 84\% CO$_2$, 16\% N$_2$, and has a surface temperature of 379 K. For a Venus analog around a 3400 K M-type star, the planet represents a 3.8 bar surface pressure consisting of 79\% CO$_2$, 21\% N$_2$, and a surface temperature of 381 K. Finally, a Venus analog around a 3000 K M type star has a surface pressure of 3.5 bar, has a 78\% CO$_2$ and 22\% N$_2$ atmosphere, and has a surface temperature of 381 K. All of these surface pressures listed are dry. The pressure-temperature profiles of each analog are represented below in Figure \ref{fig:profiles}:

\begin{figure}[!htp]
  \centering
\includegraphics[width=0.9\textwidth]{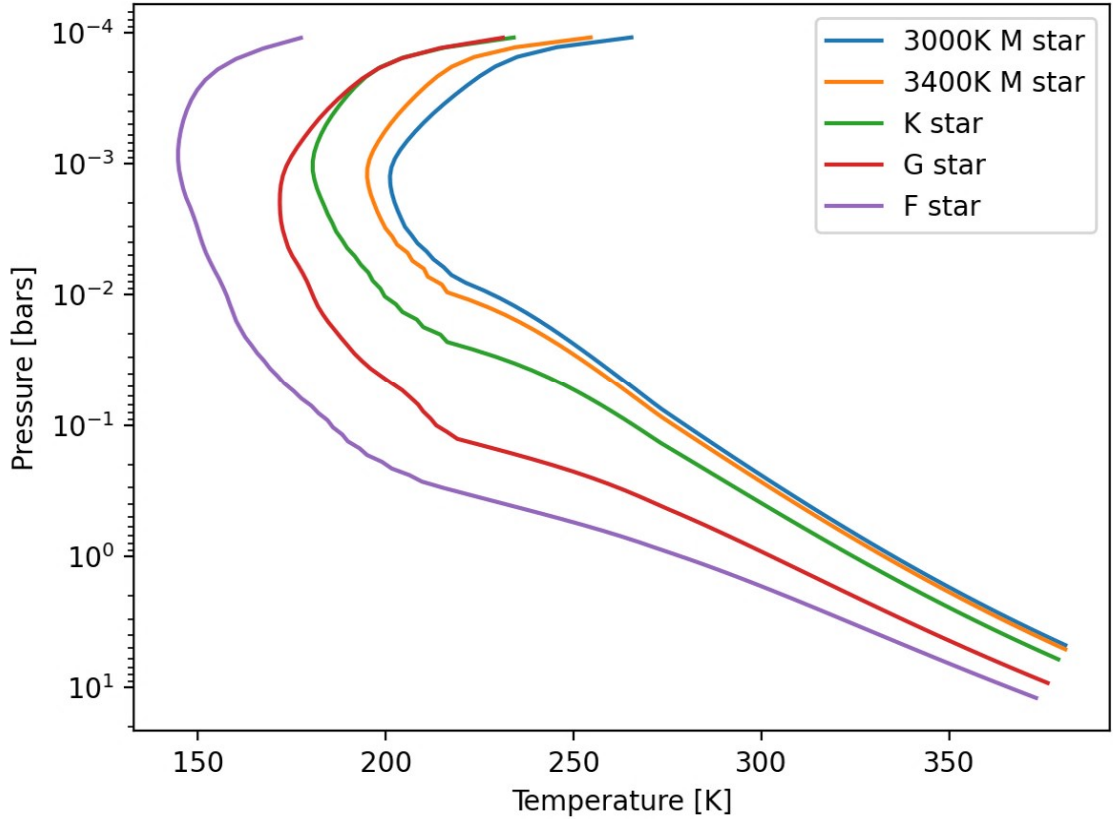}
  \caption{A comparison of pressure-temperature profiles of our Venus analogs at their last stable run for each star type, all at S\textsubscript{eff} = 1. Here, the temperature profiles for Venus analogs around F-stars tend to be cooler than the profiles for planets around M-stars. This is due to the makeup of the atmosphere in our model: the \ce{CO2} and \ce{N2} components. The F star, for example, emits peak radiation in the blue part of the spectrum than the other stars. The \ce{CO2}-\ce{N2}-\ce{H2O} atmosphere acts as an effective scattering medium for the shorter wavelength radiation, resulting in comparatively lower incoming solar radiation reaching the planet's surface and causing little warming. The M stars, on the other hand, emit their peak radiation more in the infrared part of the spectrum, which can reach the surface due to lower Rayleigh scattering. Furthermore, the Venus analog atmosphere is dominated by \ce{CO2} which is an efficient absorber of near-IR radiation resulting in a warmer profile.} 
  \label{fig:profiles}
\end{figure}
\par
Although our model does not account for photochemistry, it is worth noting that the higher amounts of UV radiation emitted by the larger, warmer stars (the F and G stars) also contribute to these P-T profile patterns. UV radiation is a major driver of atmospheric escape, resulting in low-density atmospheres. This low density means that there are less particles composing the atmosphere, which means less capacity for heat trapping and a lower amount of total motion exerted by a lower number of particles.\par
Finally, we wish to note that there are similar procedures of climate modeling of potential exo-Venuses, as well as constraining potential Venus habitability in its early years using 3D GCMs rather than 1D models \citep{kane_climate_2018, way_venusian_2020}. These 3D models provide a higher level of detail including but not limited to: the effects of clouds, wind speeds, and orbital and rotational parameters. Future investigations of the Venus zone would benefit from the use of 3D models.

\section{Results: Finding the Venus Zone} \label{sec:zone}
\par 
We use these resulting Venus analog parameters for each star type to find the distance from the star where \ce{CO2} condensation occurs. These parameters represent our incipient greenhouse conditions that can evolve into a post-runaway greenhouse. Since our aim is to probe the distance from a star in which Venus-like conditions can exist, we keep these initial analog parameters constant while decreasing the incident stellar flux relative to Earth received from the star, hereon referred to as S\textsubscript{eff}. S\textsubscript{eff} is decreased by 0.1 over each iteration, and when \ce{CO2} condenses, we find the exact S\textsubscript{eff} to two decimal places where \ce{CO2} first begins condensing. \ce{CO2} condensation is determined by the saturated mixing ratio of \ce{CO2} in the atmosphere – which is calculated by dividing the saturation vapor pressure of \ce{CO2} at each atmospheric layer by the pressure at that layer. When the calculated saturated mixing ratio of \ce{CO2} is less than or equal to the mixing ratio of \ce{CO2} prescribed for the Venus analog, condensation occurs. This is because as S\textsubscript{eff} decreases, the atmosphere becomes too cold to fully sustain the prescribed amount of \ce{CO2} in the gas phase, and \ce{CO2} would condense. Maintaining \ce{CO2} in a gas phase is critical in maintaining the high pressure atmospheres of our Venus analogs, so at this point where saturation begins occurring, our Venus-like conditions are forfeit.

The presence of \ce{CO2} condensation regions in the upper atmosphere may seem surprising given the hot climates being studied here.  However, while the \ce{CO2} greenhouse effect strongly warms the troposphere, \ce{CO2} radiatively cools the stratosphere. Thus, cold upper atmospheric temperatures are not unexpected for such worlds (e.g. \citet{wordsworth&pierrehumbert:2013}).  Thus, our deduced outer edge of the Venus-zone marks where a \ce{CO2}-dominated planet would begin to form \ce{CO2} ice clouds.  Presumably, if we were to continue reducing the stellar constant, the \ce{CO2} condensing region would grow in size, and eventually \ce{CO2} would begin to precipitate out of the planet's atmosphere at the cold traps (i.e. the poles and/or the night-side of tidally locked planets), resulting in a runaway freeze-out of the \ce{CO2} dominated atmosphere. \par

Figure \ref{fig:cond} visualizes \ce{CO2} condensation for the Venus analog around each host star type as S\textsubscript{eff} decreases.

\begin{figure}[!htp]
    \centering
    \hspace*{-0.9cm}\subfigure[F0V star, 7200 K]{\includegraphics[width = 99mm, trim=0 0 0 0]{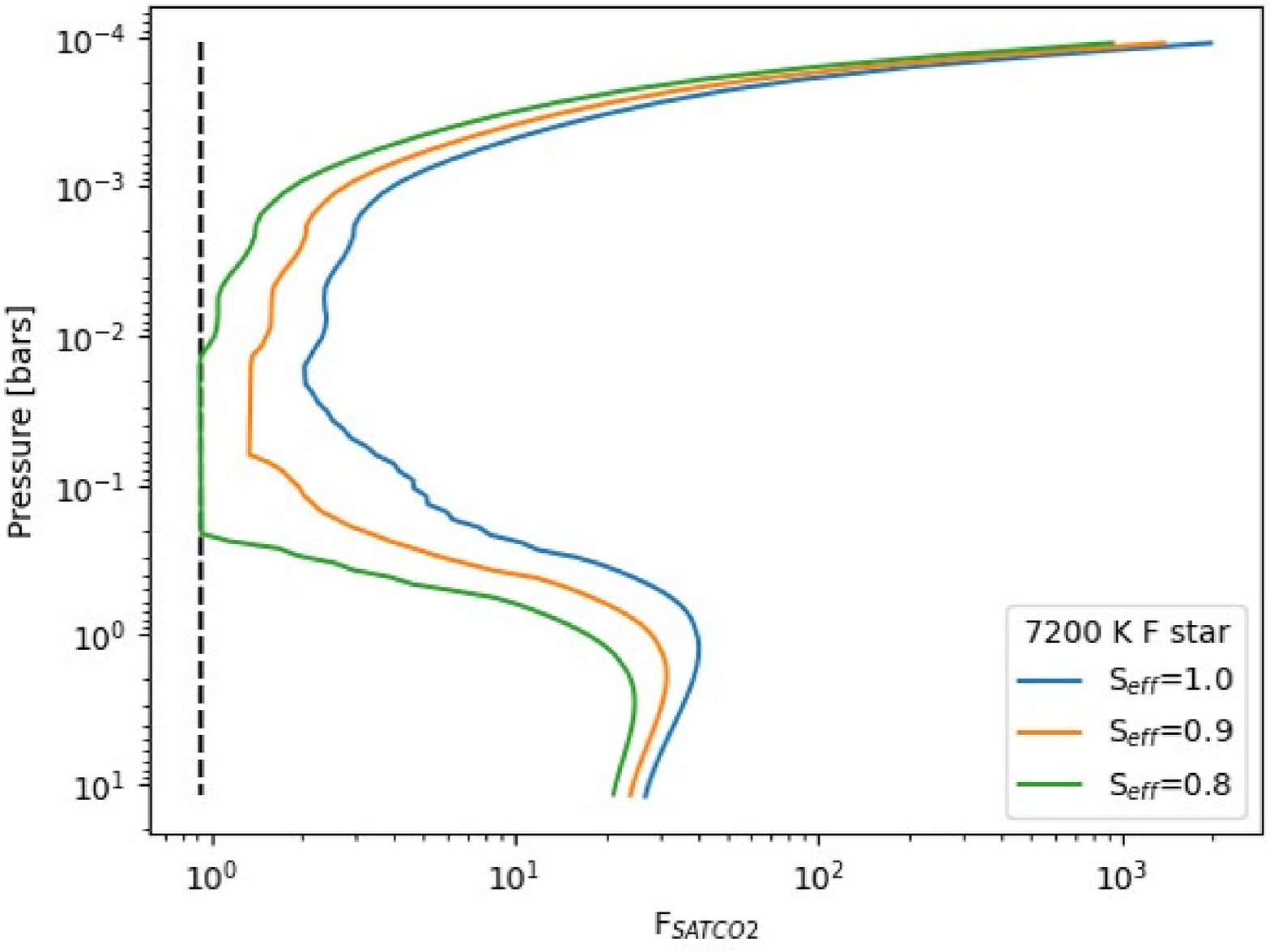}}\subfigure[G2V star, 5800 K]{\includegraphics[width = 99mm, trim=0 0 0 0]{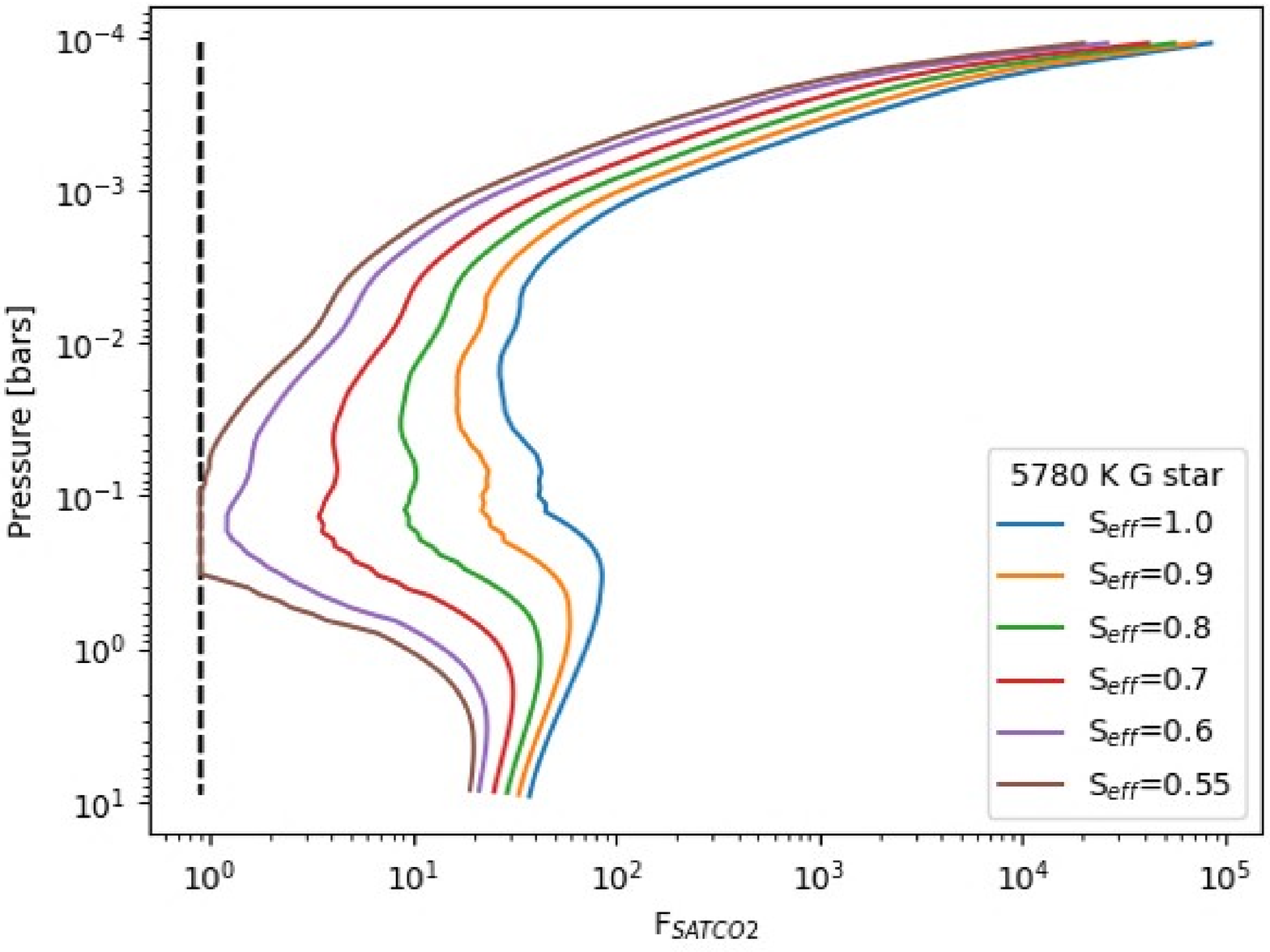}}
    
    \hspace*{-0.9cm}\subfigure[K5V star, 4400 K]{\includegraphics[width=99mm, trim=0 0 0 1cm]{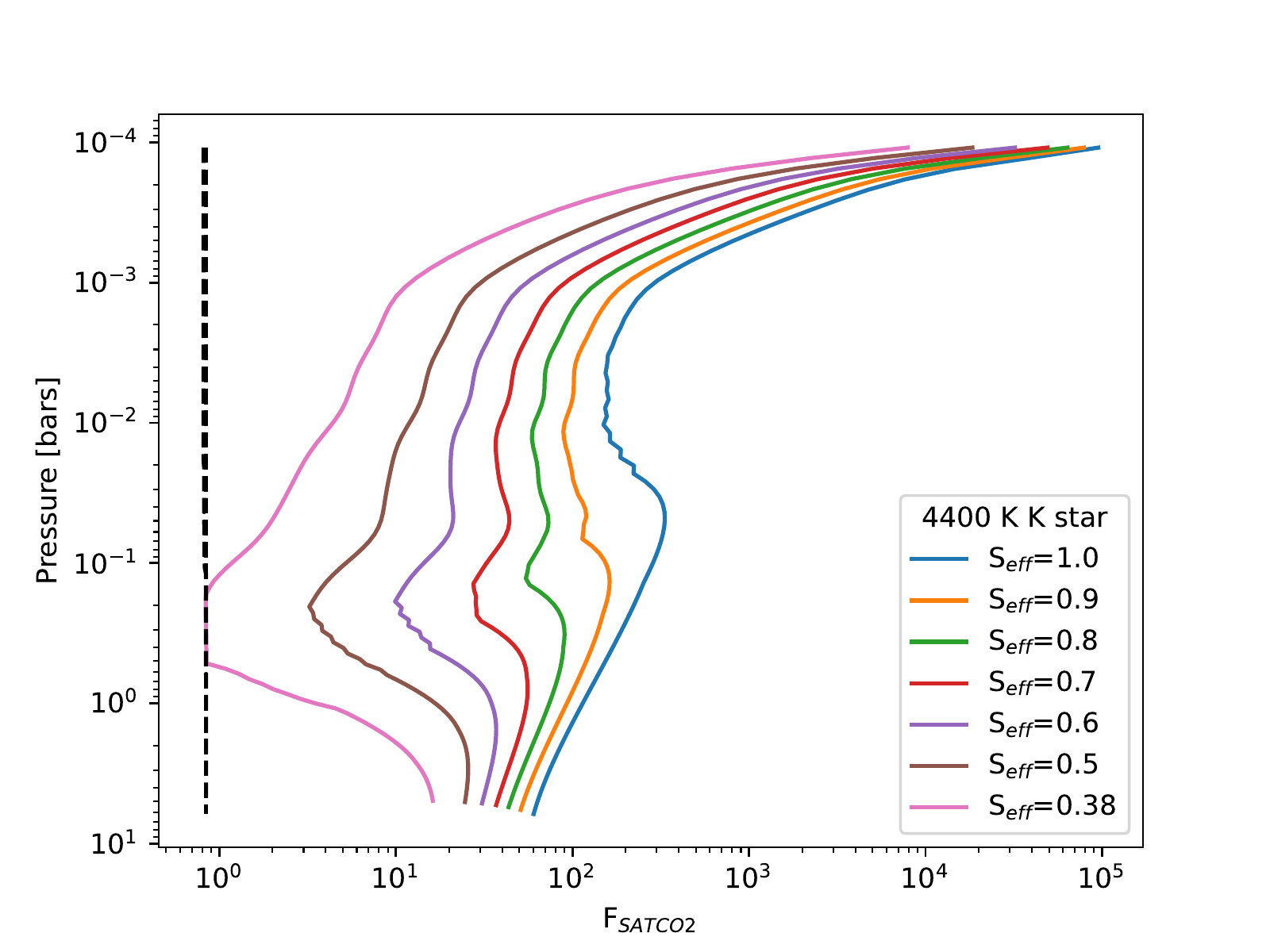}}\subfigure[M3V star, 3400 K]{\includegraphics[width = 99 mm, trim=0 0 0 1cm]{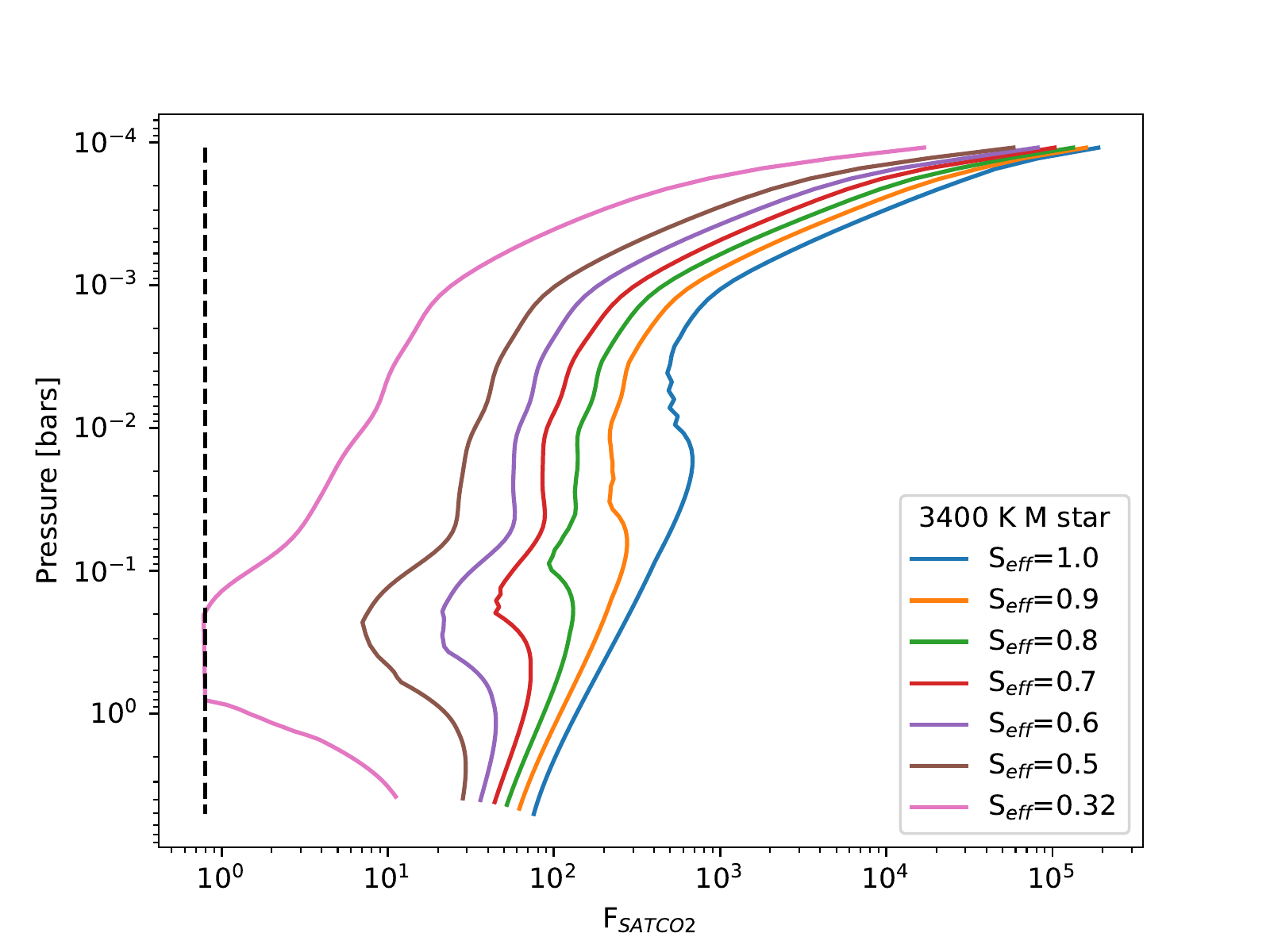}}
    
    \hspace{-0.9cm}\subfigure[M5V star, 3000 K]{\includegraphics[width = 99 mm, trim=0 0 0 1cm]{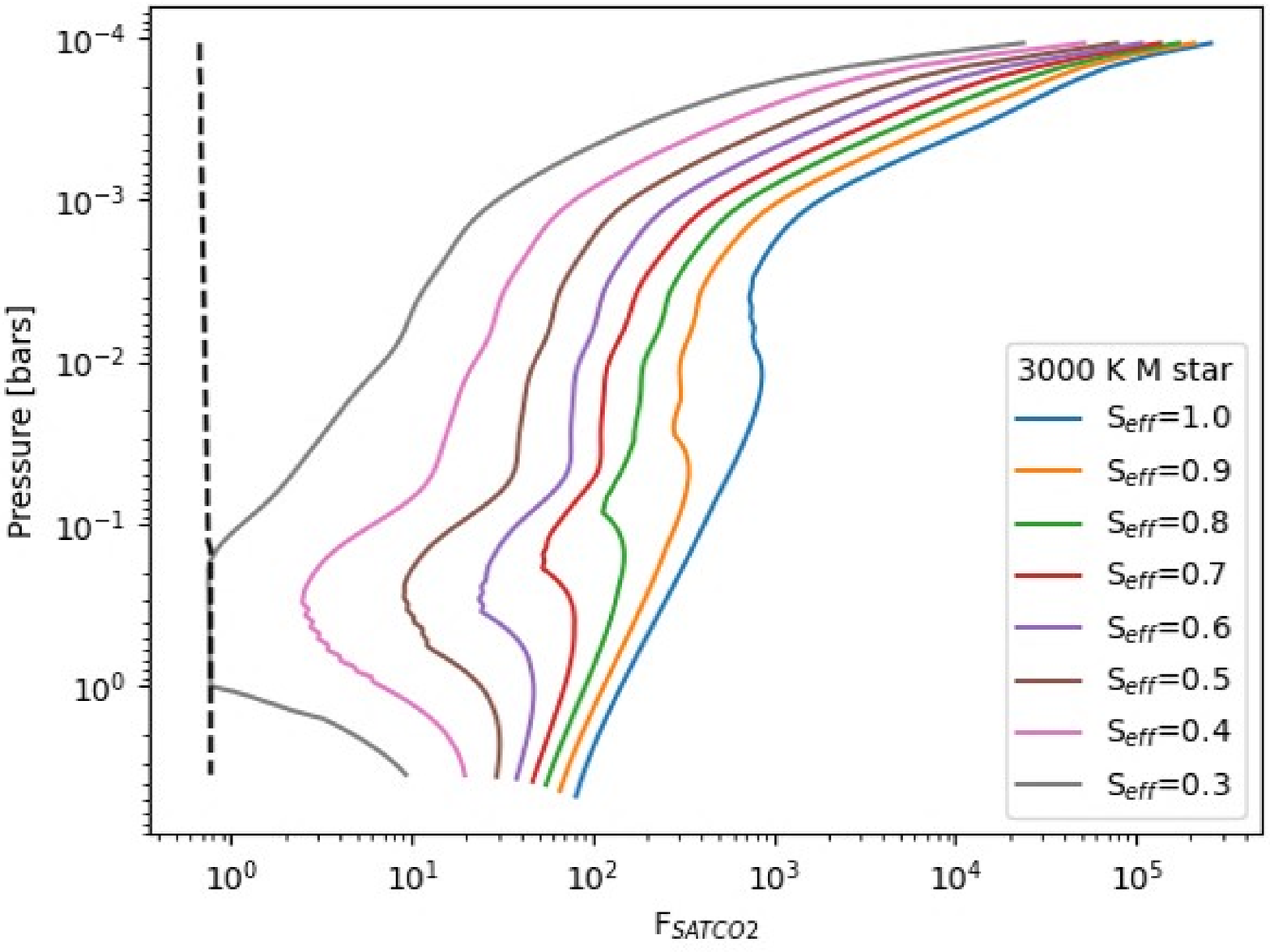}}
    %\end{figure}

\end{figure}
% \addtocounter{figure}{-1}
\begin{figure} [t!]
\caption{A comparison of \ce{CO2} condensation occurrence as a function of incident stellar flux between each star type. Each near-vertical dashed black line indicates the input mixing ratio of \ce{CO2}. The solid curves indicate the ratio of saturation vapor pressure of \ce{CO2} and the ambient pressure, labeled as $F_{\textnormal{SAT}}$. 
 Onset of \ce{CO2} condensation occurs when $F_{\textnormal{SAT}}$ converges to the \ce{CO2} volume mixing ratio; in other words, when $F_{\textnormal{SAT}}$ is less than or equal to the mixing ratio of \ce{CO2}. Note that this is equivalent to more conventional descriptions that define the onset of condensation to occur when the partial pressure of a given gas exceeds its saturation vapor pressure.
 %FSATCO$_2$ is calculated by dividing the current input of \ce{CO2} at a given layer by the pressure at that layer. 
 %When FSATCO$_2$ is less than or equal to the mixing ratio of \ce{CO2},  then \ce{CO2} is not existing in a uniform vapor state throughout the atmosphere, and has begun condensing. 
 We define the outer edge of the Venus zone as the stellar flux where \ce{CO2} begins condensing. Condensation occurs under lower amounts of incident stellar flux for the smaller, cooler stars due to the efficiency of \ce{CO2} in absorbing the higher amounts of near-IR radiation emitted by these stars, thus maintaining a warm atmosphere where \ce{CO2} is not readily condensed (See Fig. 1 temperature profile for different stars). 
 %This results in maintaining warmer temperatures under lower stellar fluxes. 
 For larger, warmer stars, condensation occurs at a higher S\textsubscript{eff} value due to scattering of blue light by atmospheric \ce{N2}, rendering these atmospheres unable to retain greenhouse warming as efficiently as a planet with \ce{CO2} around an M dwarf.}\label{fig:cond}
\end{figure}

\newpage

\par It is important to note the impact that each star type has on the resulting outer edge of the Venus zone. As shown in Figure \ref{fig:cond}, the outer edge of the Venus zone for a Sun-like G star exists at a corresponding incident stellar flux of 0.55 S\textsubscript{eff}, and for F stars, it is at an even higher stellar flux S\textsubscript{eff} of 0.8 %(the calculation for deriving distance in AU from S\textsubscript{eff} is explained later in this section).
In contrast, the outer edge of the Venus zone for a 4400 K K-type star occurs at an S\textsubscript{eff} of 0.38, and for a 3400 K M-type star, 0.32 S\textsubscript{eff}. %meaning that the Venus zone for these stars continue to exist at lower values of stellar flux received from the star. 

\begin{figure}[htp]
  \centering
\includegraphics[width=0.99\textwidth]{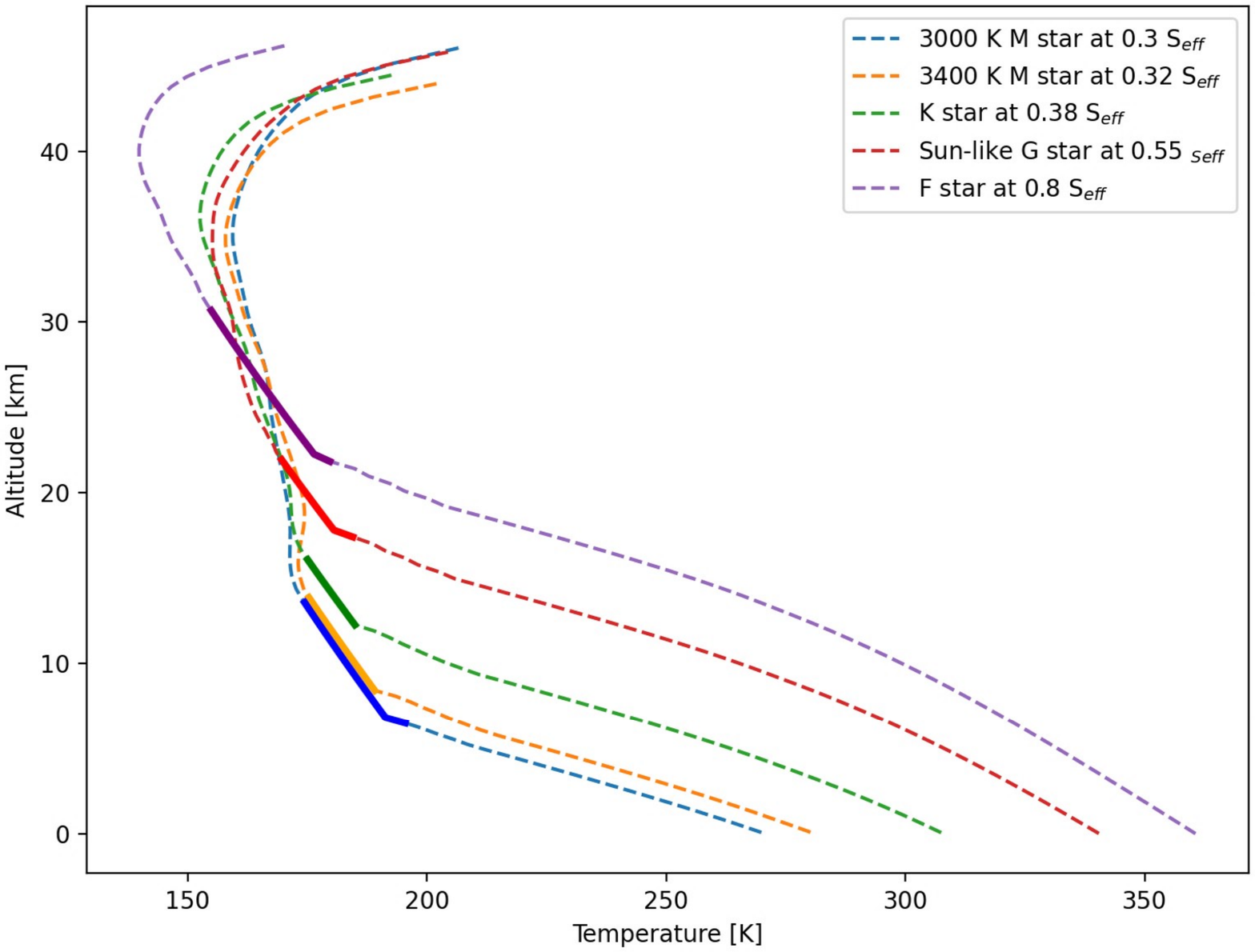}
  \caption{A comparison of \ce{CO2} condensation zones for our various Venus analogs. Solid bold bands indicate where \ce{CO2} is condensed within the atmosphere of a given Venus analog. For a 3000 K M star, \ce{CO2} begins condensing in our Venus analog at 0.3 S\textsubscript{eff}, and for a 3400 K M star, \ce{CO2} begins condensing at 0.32 S\textsubscript{eff}. For a G star, \ce{CO2} begins condensing in our Venus analog atmosphere at 0.55 S\textsubscript{eff}, whereas a Venus analog around a K star and an F star begins \ce{CO2} condensation at 0.38 and 0.8 S\textsubscript{eff}, respectively.
  %For ease of readability and to clearly distinguish the bolded regions designating condensation zones, the Y-axis of this figure is altitude, rather than pressure. 
  For M and K star profiles, \ce{CO2} condensation occurs lower in the atmosphere compared to F and G stars. As noted in Figure \ref{fig:profiles}, the \ce{N2} in our Venus analogs is effective at scattering blue light that is most present in F and G stars, while \ce{CO2} is effective at absorbing light in the infrared that is most present in M and K stars, therefore maintaining heat and additionally allowing heat to be maintained to the surface. }\label{fig:cond_compare}
\end{figure}

At equivalent S\textsubscript{eff}, the cooler, smaller stars emit more infrared and near-infrared radiation than the warmer, larger stars, which is more easily absorbed by atmospheric \ce{CO2}, allowing warming even at lower S\textsubscript{eff} values. Such warmer atmospheres can be noted in the temperature profiles shown in Figure \ref{fig:profiles}. In addition, infrared radiation is not as affected by Rayleigh scattering, as this type of scattering is wavelength dependent, resulting in more efficient scattering by shorter wavelengths (such as UV) and less efficient scattering in the IR.
%(as briefly mentioned in Figure \ref{fig:profiles})
As a result, incoming radiation from M type stars can reach lower altitudes and, with the aid of atmospheric \ce{CO2}, can be retained. Visualization of condensation in the atmosphere of each Venus analog is shown in Figure \ref{fig:cond_compare}. 

% The effects of an increase in near-infrared radiation on a CO$_2$ dominated atmosphere – rather, the efficiency of absorption of near-infrared radiation by CO$_2$ – are displayed in Figure \ref{fig:comp}.

\par Additionally, figure \ref{fig:cond_compare} shows the altitudes and temperatures where CO$_2$ can exist in a non-vapor state for each Venus analog, hereon referred to as condensation zones. These condensation zones represented by bold solid lines on the pressure-temperature profile (dashed line) of each analog at the S\textsubscript{eff} value where condensation occurs. Note that for smaller, cooler stars such as both M star and K star profiles, condensation occurs at a lower altitude than the G and F stars. In observing these planets around M and K stars, one might expect higher bond albedos than their G and F star counterparts, as lower-altitude clouds are more efficient in reflecting sunlight.\par

% \begin{figure}[H]
% \centering
% \includegraphics[width=0.99\textwidth]{0.5_compare.png}
%   \caption{A comparison between G, K, and 3400 K M type stars at 0.5 S\textsubscript{eff} shown in solid lines. 
%   %For ease of readability and to clearly distinguish each profile from one another, the Y-axis of this figure is altitude, rather than pressure. 
%   The dashed black lines indicate the region where \ce{CO2} is condensing for our G star analog. The altitude where CO$_{2}$ condenses on a planet around M and K stars are shown in dotted lines, with solid bands indicating where condensation occurs. %showing that these profiles have not yet began to condense at 0.5 S\textsubscript{eff}.
%   }\label{fig:comp}
% \end{figure}

% Figure \ref{fig:comp} emphasizes this comparison with three Venus analogs' altitude-temperature profiles as they exist at 0.5 S\textsubscript{eff}. In this case, the Sun-like G star has achieved condensation in its atmosphere, but the K and 3400 K M-type stars have not. Instead, the p-T profiles of the K and M-type stars are higher in temperature at the same S\textsubscript{eff}, displaying the efficiency of CO$_2$ in absorbing the higher near-infrared radiation emitted by K and M type stars, where a Venus zone can now extend further into these stars' habitable zones.

Finally, we can calculate the semi-major axis (distance) in AU that represents the outer edge of the Venus zone for each star type. The relationship between distance and stellar flux is represented in Equation 3 from \cite{Kopparapu2013}. The S\textsubscript{eff} is given as the ratio of incoming to outgoing radiation. So, one could multiply the S\textsubscript{eff} value that denotes each Venus zone outer edge by the solar constant value (1360 W/m2) to convert stellar flux into solar units of W/m\textsuperscript{2}. %The equation to deduct distance in AU from S\textsubscript{eff}, \textbf{giving us the outer edge of the Venus zone in AU,} is as follows:

 \begin{equation}
     \frac{d}{1 \textrm{ AU}} = \left(\frac{L/L_{\odot}}{S\textsubscript{eff}}\right) ^{0.5}
 \end{equation}
 
Where $d$ is semi-major axis, L/L$_{\odot}$ represents the luminosity of a given star relative to our Sun\footnote{Luminosity for each star type used in this paper can be found at \url{https://www.pas.rochester.edu/~emamajek/EEM_dwarf_UBVIJHK_colors_Teff.txt}}, and S\textsubscript{eff} is the stellar flux. %Stellar luminosity (in solar luminosities) for each star type is calculated as such:

%\begin{equation}
 %   L = 10^{( logL )}
%\end{equation}

Included in Table \ref{table} is the outer edge of the Venus zone for each star type, both in terms of distance in AU and in stellar flux received from the star, including the specific spectral types used in the calculation:

\par
\begin{figure}[H]
  \centering
  \hspace*{-3cm}
\includegraphics[width=1.0\textwidth]{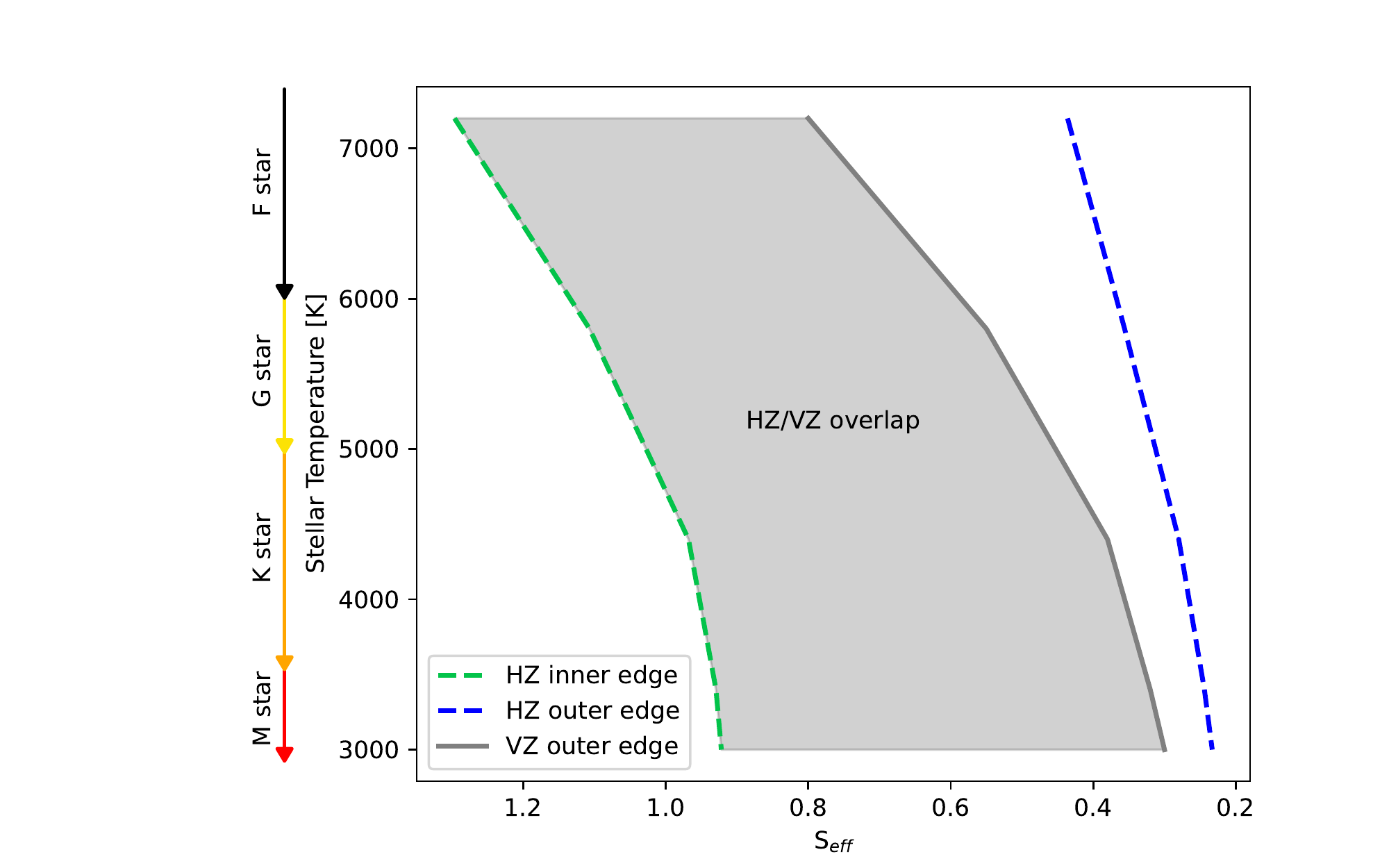}
  \hfill
  \caption{The Venus zone outer edge (gray line) as a function of stellar flux received relative to Earth, with the inner edge (green dashed) and outer edge (blue dashed) of the habitable zone included for comparison. Conservative habitable zone estimates were taken from \cite{Kopparapu2013}.}\label{fig:VZs}
\end{figure}

Figure \ref{fig:VZs} visualizes the outer edge of the Venus zone for each star type as a function of the amount of stellar flux received relative to Earth.

Table \ref{table}, in addition to our calculated values of the outer edge of the Venus Zone, also provides the conservative habitable zone estimates used in Figure \ref{fig:VZs}, both in terms of distance [AU] and incident stellar fluxed received by the planet from the star.

\begin{table}[H]
\caption{Habitable Zone and Venus Zone Boundaries}
\centering
\hspace*{-2.4cm}\begin{tabular}{ |c|c c c c c|c|}
 \hline\hline
 Star type &F0V & G2V & K5V & 3400 K M3V & 3000 K M5V \\
 \hline
 Inner HZ Distance [AU]& 2.363 & 0.950 & 0.424 & 0.132 &0.057 \\
 \hline
 Outer HZ Distance [AU] & 4.075 & 1.676 & 0.788 & 0.258 & 0.144 \\ 
 \hline
 Inner HZ [S\textsubscript{eff}] & 1.296 & 1.107 & 0.968 & 0.930 & 0.922 \\
 \hline
 Outer HZ [S\textsubscript{eff}] & 0.436 & 0.356 & 0.280 & 0.244 & 0.233 \\
 \hline
 Outer VZ Distance [AU] & 3.009 & 1.364 & 0.676 & 0.225 & 0.100 \\
 \hline
 Outer VZ [S\textsubscript{eff}] & 0.8 & 0.55 & 0.38 & 0.32 & 0.3 \\
 \hline
 \end{tabular} \label{table}
\end{table}

\section{Discussion} \label{sec:results}
We have provided calculations on the maximum possible outer edge of the Venus zone for F, G, K, and M stars using the 1D climate model \texttt{Atmos}, and updated $k$-coefficients within that model. In addition, we have identified the distance into a given star's habitable zone where we might expect to find Venus-like planets. \par

For all stars, the calculated Venus zone overlaps considerably with their respective habitable zones. For smaller, cooler stars such as M and K type stars, this Venus zone extends further into these stars' habitable zones due to their emission peaking at higher wavelengths than G type stars and the efficiency of absorption of near-infrared radiation by \ce{CO2}, as supported by Wien's law. Even for larger, brighter stars such as F and G types, their maximum Venus zones extend well into their habitable zones, as visualized in Figure \ref{fig:VZs}. There are many caveats to these results. One such caveat are our model constraints, further explained in Section \ref{sec:constraints}. Another caveat to these results is our limited understanding of geological processes that may create and support terrestrial greenhouse atmospheres. This includes, but is not limited to, the creation of these atmospheres through volcanism, cyclical surface and sub-surface processes that may prevent the evolution of these atmospheres, and an insufficient understanding of Venus' history. However, these caveats would likely impact the exact location of the outer edge of the Venus Zone or whether a specific planet would end up more like Venus or Earth or another terrestrial world. The biggest implication here - that Venus-like worlds can be present throughout much of the habitable zone - should not drastically change with updated or more nuanced models.

\subsection{Model constraints}\label{sec:constraints}
\par
The first constraint of our model to consider is the chemical constitution of our post-runaway greenhouse atmosphere used in each of our analogs. In this project, we use only \ce{CO2}, \ce{N2}, and \ce{H2O} as spectrally active species in our model, though the \texttt{Atmos} climate model can also account for CH$_4$, C$_2$H$_6$, O$_2$, \ce{O3} and, through collision induced absorption, H$_2$. It is unclear how various amounts of the other constituents supported in the model would impact our Venus zone findings. More specifically, we do not account for the presence of geological weathering processes involving these, or any, atmospheric constituents. This includes volcanic activity, responsible for the addition of sulfur dioxide and water vapor into the evolving Venusian atmosphere as well as continued surface processes involving sulfur, which aid in the propagation of Venus' modern atmosphere \citep{bullock_recent_2001}. Sulfur, and specifically sulfur dioxide, plays a large role in Venus' present day atmosphere, largely present in its clouds, and propagated by photochemistry in the middle atmosphere \citep{pinto_sulfur_2021} and thermal chemistry at the surface and subsurface, where it is believed that surface processes expedited outagssing of sulfur compounds, and specifically sulfur dioxide \citep{bullock_recent_2001}. Should these characteristics prove to be common across other greenhouse exoplanets, the development of models to adequately handle such constituents and their cycles would benefit this research. In addition, all of our Venus analogs are at 1 R$\oplus$, and it is worth exploring the existence of post-runaway greenhouse atmospheres on terrestrial planets of various sizes \citep{kasting_evolution_2003, goldblatt_low_2013, goldblatt_habitability_2015}. \par
As discussed in Section \ref{sec:atmo}, habitable zones assume inherent carbon-silicate cycling, though it is unknown how common this cycling is on terrestrial planets. Without this regulation, however, we suspect a runaway greenhouse effect would encounter minimal resistance in the form of geological processes. While it is possible that the existence of carbon-silicate cycling could have slowed Venus' evolution into a greenhouse, it is also possible that this cycling may have existed on Venus, but could not adequately prevent this evolution. Regardless of our knowledge of these cycles and their impact on these atmospheres, our model does not account for such processes, which would prove useful in understanding terrestrial atmospheres. \par

Finally, the greenhouse gas we have chosen for our Venus analogs is \ce{CO2}, as this is the dominant greenhouse gas on Venus. However, we define a Venus-like planet as a greenhouse-gas-dominated atmosphere that occurs when incoming solar radiation exceeds infrared radiation emitted at the top of a planet's atmosphere. This means that it is possible to model greenhouse atmospheres that are dominated by other greenhouse gases, such as water vapor or \ce{CH4}, using  either Atmos or other 1D climate models. Previous work has been done to examine the effect of methane on greenhouse warming in the context of the habitable zone, so examining various greenhouse gases in the context of the Venus zone could prove useful to further constrain this zone \citep{ramirez_methane_2018}.

\subsection{The post-runaway greenhouse atmosphere and implications for Earth-like planets}
\par This project visualizes the distance into a given star's habitable zone where observers may find incipient and post-runaway greenhouse atmospheres, or Venus-like planets. The capacity for the habitable zone to contain a Venus-like planet begs the understanding of how habitable planets can remain habitable; in other words, what makes Venus what it is, and how has Earth remained habitable and has not yet seen the same fate as Venus? In addition to understanding the occurrence rate of and properly modeling the carbon-silicate feedback cycle as a regulator of \ce{CO2}, as well as the inclusion of the effects of other surface processes, it is crucial to understand exactly how Venus entered its greenhouse state, and the role that magnetic field loss and the evolution of plate tectonics played in creating and maintaining its atmosphere. \par 
Current speculation of how Venus entered its greenhouse state points to liquid water on the surface evaporating as our Sun entered its main sequence, and the subsequent water vapor contributing greatly to the trapping of heat, resulting in a greenhouse \citep{kasting_response_1984, kasting_runaway_1988}. In order to understand the current state of its atmosphere, in addition to uncovering how this evolution occurred, we must look to its current surface and subsurface processes, of which little is known. For example, while Venus displays no known plate tectonic activity, it has been hypothesized that Venus once had plate tectonics, but evolved into a stagnant lid and maintained a thick lithosphere \citep{solomatov_stagnant_1996}. Another hypothesis states that Venus has maintained a stagnant lid throughout its history, and relies on high volcanic activity to maintain its atmosphere \citep{smrekar_recent_2010}. In addition, the chemical makeup and size of each of Venus' inner layers presents additional challenges to understanding its environment. A planet's inner layers are responsible for the generation of a magnetic field, which Venus does not have. While models suggest that Venus may have a similar core light metal content to Mercury, which corresponds to current mass constraints for Venus \citep{steinbrugge_challenges_2021},  the amount of light metal in the core of a Mercury analog also must imply a larger core, which is not supported by current models \citep{oneill_end-member_2021}. In contrast, a liquid inner core is most supported, bearing in mind that a fully liquid core would not support a history of a mobile lid \citep{oneill_end-member_2021}. Finally, Venus' famously slow rotation rate of 243 Earth days must also be considered, as the causes of it as well as the full extent of its impacts on Venus' environment, and specifically its subsurface processes, are also largely unknown. It has also been recently suggested that Venus did not enter its current extreme conditions, but rather maintained these conditions for its entire history, and was never habitable. This work questions whether water has ever condensed on the surface, and with a 3D global climate model, shows that water clouds forming on the night side (and maintained due to Venus' slow spin) would have had a warming effect that would have thwarted conditions for liquid water on the surface \citep{turbet_daynight_2021}. Our assumptions about Venus, and in consequence our assumptions about the evolution of habitability, operate under the condition of once-existing liquid water on the surface of Venus, which may not have been the case. Indeed, even our work on the Venus zone utilizes a 1D climate model which does not consider the effects of cloud coverage, and in this case, clouds on a slow rotating planet that, unlike Earth, would not produce dry regions that promote cooling via thermal emission on the night side \citep{leconte_increased_2013}. Because of this, regardless of rotation speed, a net warming effect may have been the norm for Venus from the beginning. In this case, although it is still theoretically possible for Earth to enter a runaway greenhouse state, Venus-like conditions may not be the natural late-stage progression of habitability itself \citep{graham_high_2021}.

In light of what is known about Venus, how has the Earth remained habitable, and what chances do planets in the habitable zone have of maintaining habitable conditions? Simple addition of \ce{CO2} into the Earth's atmosphere may be able to trigger a runaway greenhouse, as the Earth's terrestrial environment is still susceptible to a Venus-like evolution \citep{goldblatt_runaway_2012,goldblatt_low_2013}. However, Earth represents a number of features that keep this fate at bay, and though increased \ce{CO2} via anthropogenic activity poses many risks to human longevity, recent models suggest that the Earth itself would need a significant amount of \ce{CO2} added before it triggered a runaway greenhouse; addition of \ce{CO2} into the Earth's atmosphere would have much less affect on Earth as opposed to a terrestrial planet towards the inner habitable zone. \citep{doi:10.1089/ast.2014.1153, graham_high_2021}. The carbon-silicate cycling present on Earth is the more immediately apparent deterrent of a Venus-like evolution, but is not the only one. Another major factor contributing to Earth's habitability is its magnetic field, which protects its biosphere from harmful solar radiation and winds, and is powered by the cooling of its core. However, the rate at which the core cools and the age of the magnetic field itself are in conflict, referred to as the "new core paradox" \citep{doi:10.1126/science.1243477}. This conflict arises from the observed age of the Earth's magnetic field, estimated to be ~3.45 Gya, \citep{tarduno_geodynamo_2010} and the rate in which the Earth's core is cooling, ~100 K/Gya, which suggests a younger magnetic field \citep{davies_constraints_2015, lapotre_probing_2020}. The question then arises regarding the longevity of the Earth's magnetic field, whether impacts including the Moon's formation may have been involved, and whether a long-lived dynamo is unique to us, or standard in the terrestrial planets we observe. It is also worth noting that these examples of how Earth has maintained its habitability does not consider the impact of life itself on its atmosphere; specifically, the ability of element-fixing life to contribute to the atmospheric makeup of Earth. \par

It is important to note that just as a planet's presence in the habitable zone does not guarantee that it is habitable, the same can be said for planets residing within the Venus zone: these planets are not guaranteed to have a greenhouse atmosphere. This is demonstrated in the overlap of the two zones, visually represented in Figure \ref{fig:VZs}; planets in this region could have climates similar to Earth or Venus, or could even represent a completely different stable climate state that is observed on terrestrial planets \citep{goldblatt_habitability_2015}. A possible aid in the search for characteristics that point to habitable, Venusian, or other conditions, are hazes that reside on terrestrial planets within the Venus zone portion of the habitable zone, which can impact climate, and may be represented by albedo. For example, the presence of hazes on Titan and possibly the Archean Earth have decreased albedo independent of the effects of atmospheric water, in addition to providing surface cooling and shielding from UV radiation \citep{arney_pale_2017,horst_titans_2017}. Venus' hazes, composed of sulfuric acid, in addition to the planet's 96\% atmospheric \ce{CO2}, results in both a highly reflective coverage and a greenhouse effect that traps extreme temperatures. Venus' sulfuric acid is likely to be a result of volcanic activity \citep{marcq_variations_2013}. When compared with Earth's periodic volcanic haze in the upper atmosphere, the presence of water to produce storms that mitigate this haze does not exist on Venus, resulting in an optically thick haze, rendering potentially habitable planets residing within our Venus zone unlikely to represent the same albedo as Venus itself \citep{del_genio_albedos_2019}. Comparative planetology to understand common denominators of both habitable and Venus-like planets, and the distances from a given host star where we can expect to find such characteristics, would help identify signs of a post-runaway greenhouse atmosphere; perhaps our bright Venus is an outlier, or perhaps it is the norm.\par

Regardless of all the possibilities laid out above, the boundaries presented in this work are only as useful as their capacity to be tested via exoplanet observations and their atmospheric characterization. Our calculated overlap between the habitable zone and the Venus zone proposes new difficulties in testing either zone via observation. Because both the habitable zone and the Venus zone (as defined in this work) are the zones in which either climate state is possible, the overlap implies that more than one stable climate state is possible for a given stellar flux. This result is consistent with other work showing that multiple states can be stable at a given instellation \citep{goldblatt_habitability_2015}. Here, we demonstrate the spatial extent around the star for which various stable states may be present. Testing this may be possible by either statistically analyzing the abundance of habitable and Venus-like worlds in the overlap of these regions, or via refutation by finding habitable worlds beyond the habitable zone, or Venus-like worlds beyond the Venus zone.

\section{Conclusion}
Venus has long presented a challenge for scientists who seek to understand its past, present, and future. Despite a highly uninhabitable atmosphere dominated by carbon dioxide and sulfuric acid and extreme temperatures and pressure, Venus shares strikingly similar bulk composition and characteristics with Earth, and is even believed to have been habitable in the Sun's youth. Understanding Earth analogs, and subsequently understanding habitability and its evolution, requires an understanding of Venus. \par 
Upcoming DAVINCI, VERITAS, and EnVision missions to Venus will further uncover these mysteries by exploring Venus' history and answering the questions of how and when Venus transitioned into its current climate states, which will in turn further refine our understanding of the Venus zone. This will occur through these missions' explorations of the origins and evolution of Venus' atmosphere, its past and current surface processes including the rate of volcanic activity, and overall, the ways that Venus evolved differently than Earth \citep{garvin_davinci_2020, smrekar_veritas_2020}. Meanwhile, JWST will have the opportunity to explore planets in the Venus zones of nearby M-type stars \citep{gardner_james_2006}. Should there be observations of terrestrial habitable-zone planets that turn out to be Venus-like candidates, either with JWST or continued exoplanet characterization efforts, this will provide us with a better understanding of how common Venus-like planets are, and an opportunity to test the main hypothesis in this work: that Venus-like planets are able to exist well into the habitable zone of a star. \par 
With a more constrained understanding of Venusian processes and history, and observations of potentially Venus-like worlds around other stars, we can build more holistic models with a greater understanding of how our planetary twin came to be, and how the habitable worlds that we search for may have seen, or one day will see, a similar evolution. As presented in this project, when we look outside of our solar system in search of another Earth and to further understand habitability, we may just find a Venus instead. 

% \acknowledgments
% We thank Ben Hayworth for recent updates to the climate model of Atmos.

%  This work was performed by the Virtual Planetary Laboratory Team, which is a member of the NASA Nexus for Exoplanet System Science, and funded via NASA Astrobiology Program Grant 80NSSC18K0829. This work also benefited from participation in the NASA Nexus for Exoplanet Systems Science research coordination network. Monica R. Vidaurri was supported by NASA through the University of Maryland College Park under the cooperative agreement with Center for Research and Exploration in Space Science and Technology (CRESST) II, distributed to Howard University through the Award "46384-Z6121001", and also acknowledges support from the Stanford University School of Earth Dean's Graduate Scholars Fellowship. Sandra T. Bastelberger acknowledges support by NASA under award number 80GSFC21M0002. Goddard affiliates and Eric T. Wolf acknowledge support from the GSFC Sellers Exoplanet Environments Collaboration (SEEC), which is supported by NASA’s Planetary Science Division’s Research Program. 

\bibliography{bibforVenuszone}{}
\bibliographystyle{aasjournal}

\end{document}